\pdfoutput=1
\documentclass[letterpaper,10pt]{article} 
\usepackage{opticameet3} %% use version 3 for proper copyright statement

\usepackage{cite}
\usepackage{amsmath,amssymb,amsfonts}
\usepackage{algorithmic}
\usepackage{hyperref}
\usepackage{graphicx}
\usepackage[nopatch=footnote]{microtype}
\usepackage{textcomp}
\usepackage{xcolor}
\usepackage[inline]{enumitem}
\usepackage[linesnumbered,ruled,vlined]{algorithm2e}
\usepackage{svg}
\usepackage{longtable}
\usepackage{float}
\usepackage{caption}
\usepackage{acronym}
\acrodef{OLS}{Open Line System}
\acrodef{ROADM}{Reconfigurable Optical Add-Drop Multiplexer}
\acrodef{BVT}{Bandwidth Variable Transceiver}
\acrodef{OSNR}{Optical Signal to Noise Ratio}
\acrodef{ASE}{Amplified Spontaneous Emission}
\acrodef{NETCONF}{Network Configuration}
\acrodef{NLI}{Non Linear Interference}
\acrodef{BER}{Bit Error Rate}
\acrodef{OSaaS}{Optical Spectrum as a Service}
\acrodef{CFO}{Carrier Frequency Offset}
\acrodef{CDC}{Chromatic Dispersion Compensation}
\acrodef{DGD}{Differential Group Delay}
\acrodef{ORP}{Optical Received Power}
\acrodef{SNR}{Signal to Noise Ratio}
\acrodef{PDL}{Polarization Dependent Loss}
\acrodef{ML}{Machine Learning}
\acrodef{ANN}{Artificial Neural Network}
\acrodef{OCM}{Optical Channel Monitor}
\acrodef{OPM}{Optical Performance Monitoring}
\acrodef{EDFA}{Erbium Doped Fiber Amplifier}
\acrodef{EON}{Elastic Optical Network}
\acrodef{DWDM}{Dense Wavelength Division Multiplexing}
\acrodef{QoT}{Quality of Transmission}
\acrodef{WSS}{Wavelength Selective Switch}
\acrodef{OOK}{On-Off Keying}
\acrodef{PSD}{Power Spectral Density}
\acrodef{SLA}{Service Level Agreement}
\acrodef{XPM}{Cross Phase Modulation}
\acrodef{CPM}{Cross Polarization Modulation}
\acrodef{SOP}{State of Polarization}
\acrodef{PD}{Photodetector}
\acrodef{ILA}{In-Line Amplifier}
\usepackage{setspace}
\usepackage{subcaption}
\usepackage{multirow}
\usepackage[export]{adjustbox}
\usepackage{soul}

\SetCommentSty{mycommfont}

\SetKwInput{KwInput}{Input}                % Set the Input
\SetKwInput{KwOutput}{Output}              % set the Output
\SetKwFunction{KwFn}{Fn}                   % set the Function
\begin{document}
\vspace{-6mm}
\title{Real-Time Streaming Telemetry Based Detection and Mitigation of OOK and Power Interference in Multi-User OSaaS Networks}

% \author{Author name(s)}
% \address{Author affiliation and full address}
% \email{e-mail address}
%%Uncomment the following line to override copyright year from the default current year.
%\copyrightyear{2022}

\vspace{-6mm}
\author{Agastya Raj\textsuperscript{(1)},
Devika Dass\textsuperscript{(1)},
% Tian Tian\textsuperscript{(2)},
Daniel C. Kilper\textsuperscript{(1)},
Marco Ruffini\textsuperscript{(1)}}
\vspace{-1mm}
\address{\textsuperscript{(1)}CONNECT, School of Computer Science and Statistics and of Engineering, Trinity College Dublin, Ireland}
%\\\textsuperscript{(2)}Northwestern Polytechnical University, Xi'an, China}
 
\vspace{-1mm}
\email{\href{mailto:rajag@tcd.ie}{\textcolor{blue}{rajag@tcd.ie}}}

\vspace{-7mm}

\begin{abstract}
We present a framework to identify and mitigate rogue OOK signals and user-generated power interference in a multi-user Optical-Spectrum-as-a-Service network. Experimental tests on the OpenIreland-testbed achieve up to 89\% detection rate within 10 seconds of an interference event. 
\end{abstract}
\vspace{-3mm}

\section{Introduction}
\vspace{-3mm}
Recent years have seen a significant shift from fixed-grid optical networks towards more flexible and dynamic optical networking technology such as \acp{EON} and recently to \ac{OLS}. These advancements enable operators to dynamically allocate bandwidth, providing new flexible transport services such as \ac{OSaaS}~\cite{kaeval_characterization_2022}. \ac{OSaaS} allows users to lease a defined spectral window in a fiber link, operating multiple optical channels using their respective transceivers while benefiting from reduced costs and operational complexity. 
Currently, \ac{OSaaS} is primarily offered to trusted entities~\cite{geant}, such as other network operators. Extending \ac{OSaaS} to more customers would allow operators to profit from leasing excess capacity in the fibre, while users can access low-cost, high-capacity optical network resources. %One use case includes mobile operators deploying Open RAN solutions transparently across city-wide metro-access links (e.g., using Split Option 7.2 where high capacity and low latency are essential for efficient operation).

However, shared fiber access to third-party customers introduces network vulnerabilities. Inter-channel interference, such as nonlinearity and crosstalk, may arise with dense spectrum allocation, impacting the performance of co-propagating channels~\cite{pointurier_design_2017}. Additionally, if an operator has limited visibility into the customer’s allocated spectrum, it becomes challenging to identify which user's activity is causing service disruptions. This lack of visibility can also make it difficult for operators to detect rogue or misconfigured signals, such as \ac{OOK} signals in a predominantly coherent-optimized system, which may degrade the quality of neighboring channels through mechanisms like \ac{XPM}~\cite{xpm_paper, cpm_paper}. In a recent survey with 25 network operators, 60\% of operators were concerned about the power/\ac{PSD} within \ac{OSaaS}, while others were cautious of incompatible signals such as \ac{OOK} that may be injected into the system~\cite{kaida_thesis}.

To address these challenges, operators need to develop effective tools to autonomously detect and mitigate impairments caused by \ac{OSaaS} users. By targeting the impairment source, the operator can address both malicious and non-malicious user behaviors, maintaining \acp{SLA} with other users. Although the inter-channel interactions are complex, \ac{ML} can be leveraged to identify the misbehaving channels. Spectrum-blind configurations provide limited information, such as aggregate power across a spectral block, making it difficult to isolate the source of interference or ensure compliance with \acp{SLA}~\cite{osaas_ecoc_paper}. A spectrum-aware monitoring approach, although beneficial, may be costly or impractical due to excessive filtering penalties or vendor limitations on \ac{ROADM} configurations.

In this work, we propose an enhanced architecture that enables the safe operation of \ac{OSaaS} by promptly detecting and mitigating rogue \ac{OOK} signals introduced by potential customers. Experiments were carried out on a 273 km long \acf{OLS} with 4 \acp{ROADM}, where one operator provides access to three \ac{OSaaS} customers, each operating their respective edge \ac{ROADM}. Our approach employs a hybrid monitoring mechanism using fine power measurements with the internal \ac{OCM} feature. This is done by selecting adjacent \ac{WSS} windows at 6.25 GHz granularity (as explained in more detail in the next section) at the ingress \ac{ROADM} to enable fine granularity power measurements for each customer within the wider \ac{OSaaS} windows. Additionally, we leverage a \ac{ML} framework using streaming telemetry from an operator-managed probe channel to locate the rogue \ac{OOK} signal. The proposed framework is capable of detecting the rogue signals and resolving power/\ac{PSD} limit violations due to power increases or interfering ADD/DROP activities~\cite{osaas_ecoc_paper}. We achieve a detection rate of up to 89\% for \ac{OOK} signals and a complete resolution of power interferences. 
\footnote{This paper is a preprint of a paper submitted to OFC 2025.}
\vspace{-3mm}
\section{Experimental Setup}  
\vspace{-2mm}

\begin{figure}[t]
\hspace{-8mm}
\begin{minipage}[]{0.48\textwidth}

    \centering
    \includegraphics[width=0.9\linewidth]{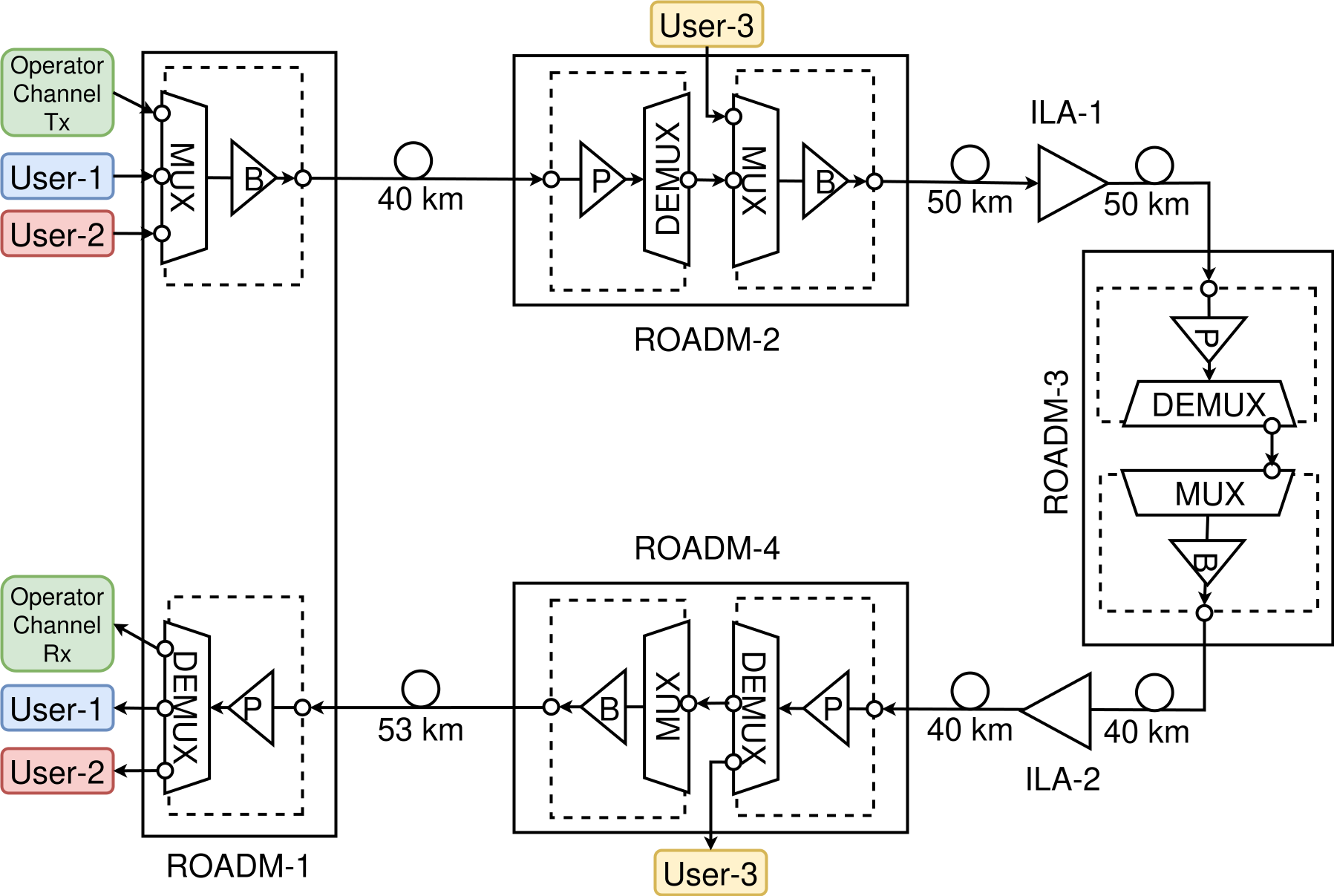}
    \caption{Experimental Setup}
    \label{topology}

\end{minipage}
\hspace{-5mm}
\begin{minipage}[]{0.48\textwidth}
  \centering
  \includegraphics[width=1\linewidth]{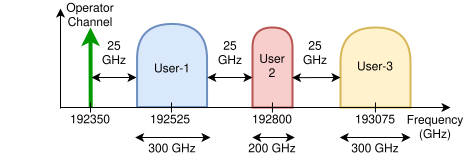}
  \vspace{-8mm}
\caption{OSaaS Channel Configuration}
\label{channel_config}

% \end{minipage}

% \begin{minipage}[]{\textwidth}
\footnotesize
    \centering
    \begin{tabular}{|c|c|c|c|}
    
    \hline
         \textbf{Parameter} & \textbf{User-1}& \textbf{User-2}&  \textbf{User-3}\\
        \hline
         % Start Freq (GHz) &  192375&  192700&  192925\\ \hline
         % End Freq (GHz) &  192675&  192900&  193225\\ \hline
         OSaaS Width (GHz) &  300&  200&  300\\ \hline
 Min/Max Power (dBm) & -2.2/11.8 & -4.0/9.8& -2.2/11.8\\ \hline
 Min/Max PSD (dBm/GHz) & -27.0/-10.8& -27.0/-10.8& -27.0/-10.8\\ \hline
 % \multirow{3}{*}{Channels configured} & TeraFlex 300 Gbps 16-QAM& TeraFlex 300 Gbps 16-QAM& TeraFlex 300 Gbps 16-QAM\\
 % & QuadFlex 200 Gbps 16-QAM& QuadFlex 200 Gbps 16-QAM& Cassini 100 Gbps DP-QPSK \\
 % &Cassini 100 Gbps DP-QPSK &Cassini 100 Gbps DP-QPSK & Cassini 100 Gbps DP-QPSK \\\hline
    \end{tabular}
    \vspace{-2mm}
    \caption{\ac{OSaaS} user configurations, and defined \acp{SLA}}
    \label{tab:my_label}

\end{minipage}

\vspace{-10mm}
\end{figure}

% Conforming to ITU-T Recommendation G.807\cite{itu}, we define an optical spectrum service as a dedicated optical media channel, facilitating a transparent lightpath through an optical network on a specific frequency slot, capable of carrying multiple optical tributary signals.
The experimental setup was implemented in the OpenIreland Testbed~\cite{open_ireland}, as shown in Fig.~\ref{topology}. The \ac{OLS} comprises four Lumentum ROADM-20 units with amplification at each node, and 2 Juniper \ac{ILA} units, configured to emulate a metro network with a total topology length of 273 km. In this study, we consider an \ac{OSaaS} scenario involving three \ac{OSaaS} end customers, as shown in Fig.~\ref{channel_config}. User-1 and User-2 are added at Node-1 and dropped at Node-5; while User-3 is added at Node-2 and dropped at Node-4. Each user is assigned a distinct spectral window across three non-overlapping network segments within the \ac{OLS}, as shown in Fig.~\ref{channel_config}.

The experiment uses a heterogeneous mix of data rates and transceiver vendors to emulate a multi-vendor, multi-user \ac{OSaaS} environment, as defined in Fig.~\ref{tab:my_label}. The \ac{PSD} and power limits have been defined according to the equipment limits for a standard 200 Gbps 16-QAM signal in a 50 GHz grid according to ITU-T G.694.1. To monitor system performance and detect rogue \ac{OOK} signals, we introduce an operator-managed Teraflex 200 Gbps 16-QAM data channel, spaced 25 GHz away from the customer spectrum. While this is a standard coherent channel used by the operator for data services, its streaming telemetry information is used to monitor nonlinear impairments in the fibre caused by the \ac{OSaaS} users.

%... first you need to explain how channel monitoring is carried out...

A key feature of this work is that by partitioning the \ac{OSaaS} spectrum into multiple adjacent 6.25 GHz \ac{WSS} channels, we can use the \ac{ROADM} \ac{OCM} to measure power at that granularity. Since the 6.25 GHz channels are adjacent, this does not cause filtering impairments if carried out only at the ingress \ac{ROADM}, as the effect is an almost flat spectrum. %Indeed, this high-granularity channelisation is only carried out at the ingress ROADM for the OSaaS users, in order to mitigate excessive filtering penalties.
%due to cascading \ac{WSS} filtering, a hybrid monitoring approach is adopted. We implement multiple adjacent 6.25 GHz \ac{WSS} media channels over the entire allotted user spectrum to configure wide-band \ac{OSaaS} at the ingress \ac{ROADM} where user's signals are added in the \ac{OLS}, providing more granular power-level monitoring. 
This also allows for precise targeted actions. For example, the operator could increase the in-\ac{WSS} Variable Optical Attenuator (VOA) value only on specific 6.25 GHz channels to block just the interfering signal within a customer's spectrum, thereby restoring services for other channels without shutting down the entire spectral block allocated to the customer creating the issue.

Streaming telemetry data is collected every second from the Teraflex probe channel using a gNMI server, which extracts performance monitoring features: Carrier Frequency Offset (CFO), Chromatic Dispersion Compensation (CDC), Differential Group Delay (DGD), optical received power, Optical Signal-to-Noise Ratio (OSNR), pre-FEC Bit Error Rate (BER), and Polarization Dependent Loss (PDL). Additionally, we collect total power measured at the ADD port, and measured power readings per configured \ac{WSS} channel in each user's ingress \ac{ROADM}. 

\vspace{-3mm}

\section{Impairments introduced by OSaaS users} 
\vspace{-2mm}
To investigate the potential impairments from the \ac{OSaaS} users, we have focused our experiments on three main use cases.
\\
\begin{enumerate*}
\item \textbf{Adding \ac{OOK} signals in the \ac{OSaaS} window}: In this case, a 10GB/s \ac{OOK} signal with 20 GHz width was introduced in user's spectrum at different frequencies. 
% Detecting rogue \ac{OOK} signals based solely on power monitoring is challenging, as \ac{OOK} signals can significantly degrade the performance of neighboring coherent channels. 
While the issues were previously investigated in~\cite{xpm_paper}, we noticed that even a slight increase in the power of an \ac{OOK} channel(+2 dBm) can lead to severe degradation of the co-propagating coherent channels, with some channels falling below the SD-FEC limit. The primary cause of this performance degradation is \ac{XPM}, which occurs due to the nonlinear Kerr effect—where the high optical intensity of the \ac{OOK} signal induces oscillations in the refractive index of the fiber material.

The impact of \ac{OOK} signals on coherent channels is most severe when they are spectrally close, with effects diminishing beyond 500 GHz of spectral separation. Additionally, \ac{OOK} channels can also cause small effects on channels with spectral separation up to 1 THz due to \ac{CPM}, resulting in sudden fluctuations in the \ac{SOP} of the affected channels~\cite{cpm_paper}. In polarization-sensitive receivers, these \ac{SOP} changes can manifest as amplitude fluctuations, which can lead to fluctuations in \ac{BER} over time. In our setup, the operator relies on the performance monitoring data from their own channel outside the OSaaS windows to detect such impairments. %, as we assume they not have direct performance monitoring capabilities within the customer \ac{OSaaS} spectral windows.
\\
  \item \textbf{Increase in power across the \ac{OSaaS} window}: In this case, the user's total spectrum power was systematically increased in 1 dB increments up to 15 dB, causing a gradual increase in impairments in the other channels.
  \\
  \item \textbf{ADD/DROP Impairment}: In this case, the total power of the user spectrum remains constant, but a number of channels within the spectrum is varied. User spectra with a fewer number of channels therefore have higher per channel power for the same overall power across the \ac{OSaaS} window. Higher channel power increases inter-channel non-linear interference.
\end{enumerate*}

%We introduced these impairments over a 5-hour service window for each user into the \ac{OLS}, and measured the end-to-end performance of the operator's probe channel and \ac{OCM} readings from the ingress \acp{ROADM}, polling every second.
 
\vspace{-4mm}
\section{Framework and Results}
\vspace{-2mm}
\begin{figure}[t]
\hspace{-8mm}
\begin{minipage}[]{0.48\textwidth}

    \centering
    \includegraphics[width=0.75\linewidth]{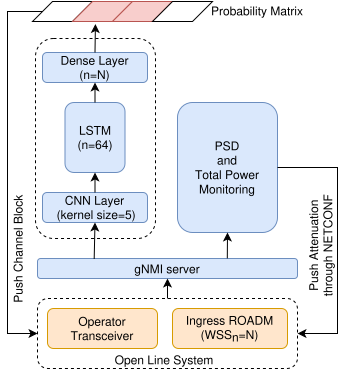}
    \vspace{-3mm}
    \caption{Modeling and Mitigation framework}
    \label{model}

\end{minipage}
\hspace{-6mm}
\begin{minipage}[]{0.48\textwidth}
    \footnotesize
    \centering
    \begin{tabular}{|c|c|c|c|c|}
    \hline
         \textbf{User}&   \textbf{Precision}&  \textbf{Recall}&  \textbf{F1 Score}& \textbf{Mean Detection Time} \\ \hline
         % &  No Interference&  100\%&  99.88\%&  99.94\%& \\ \hline
         User 1&   88.71\%&  100.0\%&  94.02\%& 2.16 s\\ \hline
         % &  No Interference&  99.94\%&  99.80\%&  99.87\%& \\ \hline
         User 2&  81.67\%&  94.04\%&  87.42\%& 6.8 s\\ \hline
         % &  No Interference&  99.80\%&  99.90\%&  99.85\%& \\ \hline
 User 3& 80.52\%& 89.00\%& 84.55\%&8.15 s\\ \hline
    \end{tabular}
    \vspace{-4.0mm}
    \caption{Performance metrics of the user-specific models for detection of OOK signal (probability threshold(p)=0.50)}
    \label{results1}

\vspace{-3.5mm}

  \centering
  \includegraphics[width=1.18\linewidth]{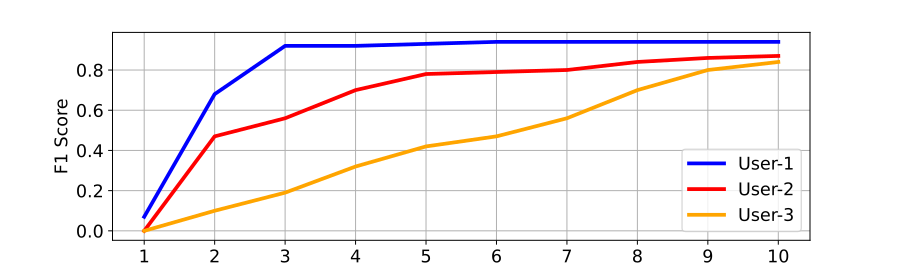}
  \vspace{-7mm}\caption{F1-Score for induced OOK impairment for each user model over time (p=0.5)}
\label{results2}

\end{minipage}

\vspace{-10mm}
\end{figure}

%In this architecture, each user has a separate modeling and mitigation framework~(Fig.~\ref{model}. 
The architecture of the proposed model is shown in Fig.~\ref{model}. The power increase is monitored using the total power at the ingress \ac{ROADM}, which has the possibility to attenuate the user's spectrum, operating the per-channel VOAs in the ROADM, when detecting a power higher than the allowed limit. \ac{PSD} is derived through OCM power measurement at 6.25GHz granularity. %Because of the granular power monitoring of the user's \ac{OSaaS} spectrum at 6.25 Ghz intervals, the total power reading and  calculation is derived, with a margin of \(\pm\)1 dBm/Ghz being added to account for the larger resolution of 6.25 GHz. 
For the ADD/DROP interference, the \ac{PSD} is calculated for each configured \ac{WSS} channel and attenuated, using the same procedure, when above the limit. 

Detecting the position of interfering \ac{OOK} signal in \ac{OSaaS} is complex, since it requires a temporal model to identify fluctuations in \ac{BER}, and a spatial model to predict the spectral location. In addition, the 6.25 GHz channel granularity for OCM measurements means the signal may partially occupy the channels at the edge of the signal.  %as well as segmenting the particular position of the interfering \ac{OOK} signal as configured in the ingress \ac{ROADM} using the \ac{OCM} power measurements. 
To address this, we use a CNN-LSTM segmentation model to predict \ac{WSS} segment containing the \ac{OOK} channel. \ac{OCM} readings from the ingress \ac{ROADM} \ac{WSS} are input in a 1D-CNN layer with a kernel size of 5, which extracts spatial features. This is concatenated with the telemetry data of the operator channel, and fed to a single-layer LSTM model with 64 hidden units. The resulting output is passed to a single dense layer with neurons corresponding to the number of \ac{WSS} channels, according to each user. The dense layer uses a sigmoid function to produce probabilities for each \ac{WSS} channel, indicating the likelihood of containing an \ac{OOK} signal. A Binary Cross Entropy loss function with Adam Optimizer (lr=1$\times$ e-03) is used to train the model over 3000 epochs. 

The trained model is deployed in real-time to analyze telemetry data from the gNMI server, with outputs directly triggering automated mitigation steps such as channel attenuation and blocking through NETCONF. To evaluate its performance, the model was tested using interferences introduced during a separate operational window. To emulate real-world scenarios, commercial 10 Gbps \ac{OOK} signals from ADTRAN metro transmission system are used at different frequencies to test the detection accuracy and latency of the framework.

Key performance metrics on the test set are presented in Fig.~\ref{results1} for each of the three users, evaluated at a model probability threshold of 50\%. In this work, we trained detection models for the three OSaaS users independently. The model achieves a 100\% recall score for user-1, with 94\% and 89\% recall scores for user-2 and user-3 respectively. The difference in performance can be attributed to the difference in spectral distance between the operator channels and the users' spectrum, which affects the severity of interference over time. Fig.~\ref{results2} shows the F1-score over the duration of the \ac{OOK} interference. The results show that user spectra near the operator channel can detect interference much faster than the users with higher spectral distance. However, all models achieve an F1-score\(>\)80\% within 10 seconds. Here, we chose a probability threshold of 50\% for immediate detection and control, but different responses can be triggered based on the model's probability. When the probability of detecting an \ac{OOK} signal exceeds a pre-defined threshold (e.g., 0.5), an alarm can be raised for alert, while automatic mitigation actions such as direct channel blocking can be performed for higher probability scores. Additionally, we achieved 100\% accuracy in mitigating power increase and ADD/DROP interferences. 
 
\vspace{-3mm}
\section{Conclusions}
\vspace{-2mm}
% \fontdimen2\font=1pt
In this work, we present a real-time framework to detect and mitigate rogue \ac{OOK} signals and power interference in a multi-user \ac{OSaaS} scenario. By leveraging a hybrid monitoring mechanism and streaming telemetry, we achieve up to 89\% recall scores in detecting rogue \ac{OOK} signals within 10 seconds, offering \ac{OSaaS} operators a potential tool to ensure service quality.

% \vspace{1.0ex}
{\footnotesize
\fontdimen2\font=1pt
\noindent\textbf{Acknowledgments.} Work supported by Science Foundation Ireland (SFI) grants 12/RC/2276 p2, 18/RI/5721, 13/RC/2077 p2 and 22/FFP-A/10598}

\end{document}